\documentclass[nameyear,MSNbibl,dvips]{arxstspdf}
\usepackage{flushend}
\usepackage{stfloats}

\volume{26}
\issue{2}
\pubyear{2011}
\firstpage{203}
\lastpage{205}
\doi{10.1214/11-STS338A}
\referstodoi{10.1214/10-STS338}

\begin{document}
\begin{frontmatter}

\title{Discussion of ``Objective Priors: An~Introduction for Frequentists'' by~M.~Ghosh}
\runtitle{Discussion}
\pdftitle{Discussion of Objective Priors: An
Introduction for Frequentists by M. Ghosh}

\begin{aug}
\author{\fnms{Jos\'{e} M.} \snm{Bernardo}\corref{}\ead[label=e1]{jose.m.bernardo@uv.es}}
\runauthor{J. M. Bernardo}

\affiliation{University of Valencia}

\address{Jos\'{e} M. Bernardo is Professor, Department of Statistics, University of Valencia, Spain \printead{e1}.}

\end{aug}



\end{frontmatter}

\section{Introduction}

Professor Ghosh has produced  a very useful, interesting piece of work
which (i) argues that Bayesian results with objective priors may be
interesting for frequentist statisticians,  (ii) reviews two useful
(unrelated) techniques which find application in the de\-rivation of
objective priors, (iii) introduces a family of divergence priors which
is claimed to include reference priors, (iv) reviews matching priors,
and (v)~demonstrates that these ideas may produce new objective
priors. I will comment in turn on each of these points.

\section{Objective Bayesian Statistics}

Professor Ghosh states that  ``with enough historical data, it is
possible to elicit a prior distribution fairly accurately.'' I believe
this is a (possibly misleading) overstatement, an example of wishful
thinking. In practice, useful prior elicitation is limited to small
text-book models with very few parameters. I have never seen a proper
elicitation job in moderately complex conventional models (say a
logistic regression), let alone in really complex problems. In optimal
circumstances, one may be able to elicit a~proper joint prior for a
couple of parameters of interest, but one is then forced to assume some
form of objective conditional prior for the many nuisance parameters
typically present in any real application. Some people  then use a
``flat'' prior, typically a~limiting form of some conjugate family of
priors; but this is a very dangerous procedure, for one does not
control the implications of the choice made, and may result in severely
biased, or even improper posteriors. There is simply no substitute for
the search of a well-motivated objective prior.

The author further states that ``Bayesian methods, if judiciously used,
can produce meaningful inferences based on$\ldots$ objective priors''
and makes reference to several problems where frequentist methods fail
to produce sensible answers, while objective Bayesian methods certainly
succeed. I surely agree with this, but I find this to be an
understatement. Ever since Wald (\citeyear{Wal50}) proved that to be
admissible (a frequentist concept!) a~procedure \textit{must} be
Bayesian, people have found, over and over again, that (as could have
been expected from this general result) the frequentist performance of
objective Bayesian procedures is typically very good, and often better
than that of the procedures derived from \textit{ad hoc} frequentist
methods. Actually, one could well invert the conventional teaching of
mathematical statistics, by teaching first objective Bayesian
methods
(motivated from first principles), and then introducing frequentist
ideas and proving that, under replication, objective Bayesian methods
\textit{also} perform very well.

\section{Asymptotic Expansions and Shrinkage}

Theorem 1 is a very useful result$\ldots$ when it is applicable. This
essentially requires conditions for the posterior to be asymptotically
normal, and we all know many important examples where this is
\textit{not} the case. It is conceivable that alternative asymptotic
expansion may similarly be obtained in those ``nonregular'' cases, and
I would like Professor Ghosh to comment on this.

$\!\!$The  shrinkage argument introduced by J. K. Ghosh was a welcome
addition to the mathematical statistician toolkit. It often provides an
elegant, efficient procedure to obtain conditional expectations. This
is another example of the power of techniques based on working on
sequences of priors based on compact sets, a procedure pioneered in the
construction of reference priors, and developed in detail in Berger,
Bernardo and Sun (\citeyear{BerBerSun09}), where these types of sequences  are used to
derive reference priors in completely general situations, with no
assumptions of asymptotic normality.

\section{Divergence priors}

Professor Ghosh recalls that in the original paper on reference priors
(Bernardo, \citeyear{Ber79}), these are obtained by (heuristically) maximizing the
\textit{expected} KL divergence (better known as Shannon expected
information), as the number of replications goes to infinity, and
quotes a later result---actually published in Berger, Bernardo and
Mendoza (\citeyear{BerBerMen89}), not in Berger and Bernardo (\citeyear{BerBer89}),
where it is proven that maximization for a finite $n$ may lead to a
discrete prior. It may be worth it to point out  that   in reference
analysis one does \textit{not} let the sample size $n$ go to infinity,
but consider  $k$ replications of the original   experiment and let $k$
go to infinity, which may be \textit{ very} different. Indeed, as a
direct consequence of this, the reference prior may depend on the
design (two sample problems provide many examples of this situation;
 see Bernardo and P\'{e}rez (\citeyear{BerPer07}) on the comparison of normal means for a
 relatively recent example). How is this implemented using the expansion techniques?

Moreover, although the mathematical consequen\-ces are very nice, the
original reason to consider an infinite amount of replications was
\textit{not} mathematical convenience, but first principles: one wants
to find the prior that maximizes the \textit{missing information} about
the quantity of interest, and the complete missing information would
only be attained by an infinite number of replications.

The fact that (with only  one parameter and under regularity conditions
which guarantee asymptotic normality) the missing information is
maximized by Jeffreys' prior for \textit{all} the information measures
derived from a family of divergences which encompass both KL and
Hellinger  is reassuring, in that the result seems to be pretty stable
with respect to changes in the definition of information. That said, we
would argue that there are many independent arguments (additivity, for
one) suggesting that Shannon is \textit{the} appropriate measure of
information in mathematical statistics. It follows that I am very
suspicious of the properties of the priors derived by maximizing the
expected chi-squared distance.  In particular, in the binomial case, I
fail to see any reason to prefer a $\operatorname{Beta} (1/4, 1/4)$ prior over Jeffreys'
$\operatorname{Beta} (1/2, 1/2)$ well justified from many (really many!) points of view.
May the author provide any such reason?

The concept of general divergence priors conceptually includes that of
reference priors in that it uses a~family of expected divergences which
includes Shannon as a particular, limiting case. The specifics of the
paper, however, exclusively refer to the relatively simple situation
where asymptotic normality may be guaranteed. I would like to see some
examples of  ``nonregular'' problems solved before concluding that the
techniques described in this paper may be used in general. Nonregular
problems were already solved in the original (Bernardo, \citeyear{Ber79})
formulation of reference priors, and have been rigorously analyzed in
Berger, Bernardo and Sun (\citeyear{BerBerSun09}).

\section{Probability matching}

As mentioned above, it is certainly interesting to analyze the
frequentist properties of objective Baye\-sian results, but one does not
necessarily want to reproduce frequentist behavior. For instance, in
the ratio of normal means problem mentioned by Professor Ghosh in the
Introduction, one certainly does \textit{not} want to ``match'' the
unacceptable coverage properties of the conventional frequentist
solutions.

More importantly, I see invariance under reparameterization as a
\textit{necessary} prerequisite for any general procedure to derive
objective priors, for the resulting  (presumably objective) inferences
cannot possibly depend on the arbitrary (and hence irrelevant)
parameterization chosen to formalize the problem. It follows that,
although it is certainly useful and important to study the eventual
matching properties of priors, I believe that requiring matching is
\textit{not} a sensible procedure to choose an objective prior.

\section{New objective priors}

Professor Ghosh states ``I believe very strongly that many new priors
will be found in the future by either a~direct application or slight
modification of these tools,''  and I agree that this is indeed quite
plausible. However, a \textit{new} objective prior is not something
necessarily a good prior. One needs objective priors which satisfy a
number of desiderata: general applicability, appropriate
marginalization properties, invariance, strong consistency, and so on
[see Bernardo (\citeyear{Ber05}), for a general discussion]. And new priors which
do not satisfy those desiderata should probably not even be considered.
Some would say that  ``the proof of the pudding is in the eating'';
that may be so, but then I would like Professor Ghosh to quote at least
one convincing example where he would propose to use an objective prior
which is \textit{not} a reference prior.


\end{document}